\documentclass[12pt]{article}
\usepackage{color}
\usepackage{amsfonts,amssymb,amsmath}
\usepackage{theorem,cite}
\newtheorem{definition}{Definition}[section]

\newtheorem{theorem}[definition]{Theorem}
\newtheorem{corollary}[definition]{Corollary}
\newtheorem{lemma}[definition]{Lemma}

\theorembodyfont{\rmfamily}
\newtheorem{rmk}{Remark}[section]

\numberwithin{equation}{section}

\definecolor{brique}{rgb}{.9,.2,0}
\definecolor{blvert}{rgb}{0,.8,.85}
\definecolor{vertcl}{rgb}{0,1,.7}

\newcommand{\bea}{\begin{eqnarray}}
\newcommand{\eea}{\end{eqnarray}}
\newcommand{\beano}{\begin{eqnarray*}}
\newcommand{\eeano}{\end{eqnarray*}}
\newcommand{\beq}{\begin{equation}}
\newcommand{\eeq}{\end{equation}}
\newcommand{\nonu}{\nonumber \\}

\newcommand{\hs}[1]{\hspace{#1 mm}}

\newcommand{\eps}{\epsilon}

\newcommand{\vph}{\varphi}

    \def\cH{{\cal H}}

\def\cP{{\cal P}}        
    \def\cT{{\cal T}}    
\def\cV{{\cal V}}

\def\fh{{\mathfrak h}}

\newcommand{\CC}{{\mathbb C}}

\newcommand{\II}{{\mathbb I}}

\newcommand{\KK}{{\mathbb K}}

\newcommand{\ZZ}{{\mathbb Z}}

\newcommand{\prt}{\partial}

\newcommand{\wh}[1]{\widehat{#1}}
\newcommand{\wt}[1]{\widetilde{#1}}
\newcommand{\wchk}[1]{\stackrel{\;\raisebox{-.1ex}{\mbox{\tiny$\vee$}}}{\raisebox{.1ex}{#1}}\!}
\newcommand{\mb}[1]{\hs{4}\mbox{#1}\hs{4}}

\newcommand{\half}{\frac{1}{2}}

\newcommand{\ket}[1]{| #1 \rangle}

\begin{document}
\renewcommand{\thefootnote}{\arabic{footnote}}
\setcounter{footnote}{0}
\newpage
\setcounter{page}{0}

\pagestyle{empty}

\null
\vfill
\begin{center}

\begin{center}
{\Huge \textsf{A new dynamical reflection algebra\\[1.ex]
 and\\[1.2ex] 
 related quantum integrable systems}}
\vspace{2cm}

{\large  \textbf{J. Avan}$^{a}$ and \textbf{E. Ragoucy}$^{b}$ }

\vspace{10mm}

\emph{$^a$ \textbf{LPTM}, CNRS and Universit\'e de Cergy-Pontoise,\\[.242cm]
2 avenue A. Chauvin, 95302 Cergy-Pontoise Cedex, France
}
\\
E-mail: avan@u-cergy.fr\\

\vfill

\emph{$^b$ \textbf{LAPTh}, 
CNRS and Universit{\'e} de Savoie,\\[.242cm]
9 chemin de Bellevue, BP 110, 74941, Annecy-Le-Vieux Cedex, 
France}
\\
E-mail: ragoucy@lapp.in2p3.fr\\

\vfill\vfill

\begin{abstract}
We propose a new dynamical reflection algebra, distinct from the 
previous dynamical boundary algebra and semi-dynamical reflection 
algebra. The associated Yang-Baxter equations, coactions, fusions, and 
commuting traces are derived. Explicit examples are given and quantum 
integrable Hamiltonians are constructed. They exhibit features similar 
to the Ruijsenaars-Schneider Hamiltonians.
\end{abstract}
\end{center}

\vfill

\end{center}

\vfill
\centerline{MSC: 81R12, 16T15, 16T25}
\centerline{Keywords: Integrable systems; Dynamical algebras; Reflection algebras}
\vfill

\rightline{June 2011}
\rightline{LAPTH-019/2011}
\rightline{\texttt{arXiv:1106.3264 [math-ph]}}
\rightline{hal-00601317}

\baselineskip=16pt

\newpage
\pagestyle{plain}

\section{Introduction\label{sec:intro}}
Quantum reflection algebras originated from the studies of 
consistent sufficient conditions on boundary processes required to 
preserve quantum integrability for systems with boundaries. 
Originating with the study of systems on a half-line \cite{che},
they were systematically developed by Sklyanin \cite{skly} (see 
also \cite{Ku1}) and more generally extended to the form of quadratic 
exchange algebra in \cite{MaFrei1}. The general structure is then:
\beq
A_{12}\,K_{1}\,B_{12}\,K_{2}\, = \, K_{2}\,C_{12}\,K_{1}\,D_{12}
\label{eq:ABCD}
\,,
\eeq
where $K$ is the matrix encapsulating the generators of the quadratic 
algebra and $A$, $B$, $C$ and $D$ are $\CC$-number structure matrices. 
Consistency of (\ref{eq:ABCD}) with requirements that 
the quantum exchange algebra be associative is guaranteed by the Yang-Baxter equation for 
$A$ and $D$, and adjoint Yang-Baxter for $A$, $B$ and for $C$, $D$. 
In particular, $A$ and $D$ are thus identified with quantum 
$R$-matrices.

Here and throughout the paper, the indices $1,2,..$ in $ABCD$ and $K$ 
label the so-called "auxiliary spaces", i.e. a vector space $\cV$ on 
which the representation in (\ref{eq:ABCD}) is defined. The space 
$\cV$ may be finite or infinite dimensional: this latter includes the 
case when $\cV$ is a loop space $V\otimes\CC[[z]]$ and $A,B,C,D$ and 
$K$ then depend on (respectively) two or one complex spectral parameter(s). Explicit dependence in 
spectral parameters may be omitted to lighten notation, except when
shifts (in the spectral parameter) occur: in that case, it will be 
explicitly written, as e.g. in section \ref{sec:expreal}. Note however 
that the dynamical parameter $\boldsymbol q$ may be explicit, even when the 
spectral parameter is omitted and attached to the auxiliary space index.
 Hence $A_{12}$ stands for
$A_{12}(\boldsymbol q)$ that is itself understood as $A_{12}(z_{1},z_{2};\boldsymbol q)$,
while $A_{21}\equiv A_{21}(\boldsymbol q)\equiv A_{21}(z_{2},z_{1};\boldsymbol q)$.

The possibility of extending the notion of reflection algebra to the 
case where $A$ and $D$ become dynamical quantum $R$-matrices arose 
first when defining integrable boundary conditions for IRF models 
\cite{BPOB}. The "boundary dynamical algebra" there defined reads:
\beq
A_{12}(\boldsymbol q)\,K_{1}(\boldsymbol q+h^{(2)})\,B_{12}(\boldsymbol q)\,K_{2}(\boldsymbol q+h^{(1)})
\, = \, 
K_{2}(\boldsymbol q+h^{(1)})\,C_{12}(\boldsymbol q)\,K_{1}(\boldsymbol q+h^{(2)})\,D_{12}(\boldsymbol q)
\label{eq:ABCDdyn}
\,,
\eeq
where $\boldsymbol q$ encodes the dynamical parameters, interpreted as 
coordinates $q_{i}$, $i=1,\ldots,r$, on the dual $\fh^*$ of the Cartan subalgebra $\fh$ of some 
Lie algebra.
$h^{(j)}_{i}\in\fh$, $i=1,\ldots,r$, are its
Cartan generators acting in the $j^{th}$ space, and the spectral parameter dependence has been 
omitted (see above). 
The expressions of type $K_{1}(\boldsymbol q+h^{(2)})$ are understood 
once one assumes that each auxiliary space $\cV$ is a diagonalizable 
module of $\fh$. In this case, $K_{1}(\boldsymbol q+h^{(2)})$ acts on 
a basis vector of $\cV_{1}\otimes\cV_{2}$ as
\beq
K_{1}(\boldsymbol q+h^{(2)})\ \ket{v_{1}}\otimes\ket{v_{2}} =
\Big(
K_{1}(q_{1}+\mu\,\lambda_{1}(v_{2}),\ldots,q_{r}+\mu\,\lambda_{r}(v_{2})) \ket{v_{1}}\Big)\otimes\ket{v_{2}} 
\eeq
where $h_{i}\,\ket{v_{2}}= \lambda_{i}(v_{2})\,\ket{v_{2}}\,,\ \forall 
i=1,\ldots,r$, and $\mu$ 
is an overall shift scale which shall be kept throughout the paper.

$K_{1}(\boldsymbol q+h^{(2)})$ can be rewritten consistently as
\beq
K_{1}(\boldsymbol q+h^{(2)})= \exp\Big(\sum_{i=1}^r 
\mu\,h_{i}^{(2)}\,\prt_{q_{i}}\Big)\, K_{1}(\boldsymbol q)\,
\exp\Big(-\sum_{i=1}^r 
\mu\,h_{i}^{(2)}\,\prt_{q_{i}}\Big)
=e^{h^{(2)}\prt}\, K_{1}(\boldsymbol q)\,e^{-h^{(2)}\prt}
\eeq

$A_{12}$ is the dynamical Yang-Baxter $R$-matrix defining 
the bulk IRF model, and obeying the dynamical Yang-Baxter or 
Gervais-Neveu-Felder equation \cite{GeNe,Fe}:
\beq
A_{12}(\boldsymbol q+h^{(3)})\,A_{13}(\boldsymbol q)\,A_{23}(\boldsymbol q+h^{(1)})
\, = \,A_{23}(\boldsymbol q)\,A_{13}(\boldsymbol q+h^{(2)})\,A_{12}(\boldsymbol q)
\label{eq:dynYBE}
\,,
\eeq
In the original case \cite{BPOB} one had $A_{12}=B_{21}=C_{12}=D_{21}$.
Exchange algebra (\ref{eq:ABCDdyn}) can however be extended to a general 
situation with unrelated $A$, $B_{12}=C_{21}$ and $D$ obeying a 
coupled set of dynamical Yang-Baxter equations together with suitable 
zero-weight and unitary relations. In this way, one gets 
a first dynamical extension of the quadratic exchange algebra 
(\ref{eq:ABCD}) defined e.g. in \cite{Kor1, NA} as "boundary dynamical" 
of "fully dynamical" quadratic exchange algebra.

A second inequivalent dynamical extension arose \cite{ACF1,ACF2} when 
studying the Poisson structure of the Lax formulation for the 
Ruijsenaars-Schneider model \cite{RS}. It reads
\beq
A_{12}(\boldsymbol q)\,K_{1}(\boldsymbol q+h^{(2)})\,B_{12}(\boldsymbol q)\,K_{2}(\boldsymbol q)\,
\, = \, K_{2}(\boldsymbol q+h^{(1)})\,C_{12}(\boldsymbol q)\,K_{1}(\boldsymbol q)\,D_{12}(\boldsymbol q)\,
\label{eq:RS}
\,.
\eeq
Studied in \cite{NAR}, it was shown in \cite{AR1} to be a deformation 
by a Drinfeld twist of a non-dynamical bulk algebra $RTT=TTR$, 
following the ideas developed in \cite{JKOS,ABRR,BRT2,BRT},  albeit 
with an associated non-trivial dynamical commuting quantum trace 
\cite{NADR} (see also \cite{AZ}).

We propose in this paper a third, a priori inequivalent and yet 
unknown, dynamical extension of the quantum exchange algebra. It reads
\beq
A_{12}(\boldsymbol q)\,K_{1}(\boldsymbol q-h^{(2)})\,B_{12}(\boldsymbol q)\,K_{2}(\boldsymbol q+h^{(1)})
\, = \, K_{2}(\boldsymbol q-h^{(1)})\,C_{12}(\boldsymbol q)\,K_{1}(\boldsymbol q+h^{(2)})\,
D_{12}(\boldsymbol q)\,
 \label{eq:ABCDdyn3}
\,.
\eeq
Its particular, alternate-sign structure was suggested to us by 
preliminary studies \cite{AR2} of the second Poisson structure of the 
Calogero-Moser rational model \cite{BF1,Ma1}. 

We shall now describe 
the key features of this third dynamical "reflection" algebra. We first 
explain its occurrence from considerations on classical quadratic 
Poisson studies (section \ref{sec:classdyn}) and their associated dynamical 
Yang-Baxter equations. We then give in section \ref{sec:expreal} explicit 
realizations of its structure matrices and generating matrix $K$
stemming from a bilinear formulation in terms of a quantum bulk 
T-matrix and its transposed. This allows a characterization of 
(\ref{eq:ABCDdyn3}) as a general "dynamical twisted Yangian" structure.
We establish a dynamical commuting quantum trace formula in section 
\ref{sec:commuting}. We derive the fusion 
structures and dressing procedures of the dynamical reflection algebra in sections \ref{sec:fusion} and \ref{sec:dressing} respectively. 
We finally treat a simple
 example of dynamical reflection algebra, yielding explicit quantum integrable Hamiltonians through application
of the trace construction..

\section{From classical to quantum quadratic exchange algebra\label{sec:classdyn}}
Our proposition for a third quantum dynamical exchange algebra 
(generalized reflection algebra) follows from our observation of a 
particular form of the classical Yang-Baxter equations obeyed by some 
structure matrices in the description of the second Poisson structure 
of the rational Calogero-Moser model. It is therefore instructive to 
explain the connection between quantum dynamical reflection algebras 
and classical dynamical Yang-Baxter equations for general quadratic 
Poisson structures. 

We shall from now on restrict ourselves to dynamical YB equations 
for which the defining Cartan algebra $\fh$ is the Cartan algebra
of $A_{n-1}$. 

Let us first of all consider the classical structures. Given a 
classical Lax matrix $\ell$, the most general quadratic form for the 
associated Poisson structure is
\beq
\{ \ell_{1}\,,\, \ell_{2}\} = a_{12}\, \ell_{1}\,\ell_{2}+ 
\ell_{1}\,b_{12}\,\ell_{2}-\ell_{2}\,c_{12}\,\ell_{1}
-\ell_{1}\,\ell_{2}\,d_{12}
\label{eq:dyn-class}
\eeq
where consistency conditions imply that $a_{12}=-a_{21}$, 
$d_{12}=-d_{21}$, $b_{12}=c_{21}$.
Note that (\ref{eq:dyn-class}) implies that the functions 
$\{\mbox{tr}\,\ell^m\,,\ m\in\ZZ_{+}\}$ Poisson-commute if $a+b=c+d$. A 
more general trace formula, 
$\mbox{tr}\,(\gamma^{-1}\,\ell)^m$, occurs whenever a scalar matrix $\gamma$ 
exists such that
\beq
a_{12}\, \gamma_{1}\,\gamma_{2}+ 
\gamma_{1}\,b_{12}\,\gamma_{2}-\gamma_{2}\,c_{12}\,\gamma_{1}
-\gamma_{1}\,\gamma_{2}\,d_{12}
=0\,,
\eeq
see \cite{MaFrei1}. 

Dynamical dependence of $abcd$ now is assumed to be on coordinates 
$q_{i}$, $i=1,\ldots,n$, on a dual $\fh^*$ of the Cartan subalgebra 
$\fh$ in $sl(n,\CC)$.

Associativity for the PB structure (\ref{eq:dyn-class}) is implied by 
algebraic consistency conditions (Yang-Baxter classical equations) 
for $a$, $b$, $c$, $d$, provided the a priori undetermined bracket 
$\{r_{12}\,,\,\ell_{3}\}$, $r=a,b,c,d$, be of an algebraic form. We 
consider here the following form for this PB
\bea
\{r_{12}\,,\,\ell_{3}\} &=& \eps_{R}\,(h_{3} \prt\, r_{12})\,\ell_{3}+
\eps_{L}\,\ell_{3}\,h_{3} \prt\, r_{12} 
\label{eq:rl-class}\\
h\prt &=& \sum_{i=1}^n \mu\,e_{ii}\otimes \frac\prt{\prt q_{i}}\,,
\label{eq:h-prt}
\eea
where $e_{ii}\in\fh$, $\eps_{R},\eps_{L}$ are c-numbers to be later determined.

Jacobi identity for the PB is ensured by the following (sufficient) 
classical dynamical Yang-Baxter equations:
\bea
{[a_{12}\,,\,a_{13}]}+{[a_{12}\,,\,a_{23}]}+{[a_{32}\,,\,a_{13}]}+
\eps_{R}\Big( h_{3}\prt\, a_{12}+h_{1}\prt\, a_{23}+h_{2}\prt\, a_{31}\Big)
=0 \,, \label{eq:class-dYBE-a}\\
{[d_{12}\,,\,d_{13}]}+{[d_{12}\,,\,d_{23}]}+{[d_{32}\,,\,d_{13}]}+
\eps_{L}\Big( h_{3}\prt\, d_{12}+h_{1}\prt\, d_{23}+h_{2}\prt\, d_{31}\Big)
=0 \,, \label{eq:class-dYBE-b}\\
{[a_{12}\,,\,c_{13}+c_{23}]}+{[c_{13}\,,\,c_{23}]}-
\eps_{L}\,h_{3}\prt\, a_{12}+\eps_{R}\,h_{1}\prt\, c_{23}-\eps_{R}\,h_{2}\prt 
\,c_{13}
=0 \,,\label{eq:class-dYBE-c} \\
{[d_{12}\,,\,b_{13}+b_{23}]}+{[b_{13}\,,\,b_{23}]}-
\eps_{R}\,h_{3}\prt\, d_{12}+\eps_{L}\,h_{1}\prt\, b_{23}-\eps_{L}\,h_{2}\prt 
\,b_{13}
=0 \,.\label{eq:class-dYBE-d}
\eea
In absence of dynamical term, one recovers the usual classical 
quadratic algebra \cite{MaFrei1}.

One then immediately observes that 
(\ref{eq:class-dYBE-a})-(\ref{eq:class-dYBE-d}) is a classical 
limit ($\hbar\to0$) of a set of 4 dynamical Yang-Baxter equations:
\bea
A_{12}(\boldsymbol q)\,A_{13}(\boldsymbol q-\eps_{R}\,h^{(2)})\,A_{23}(\boldsymbol q)
\, = \,A_{23}(\boldsymbol q-\eps_{R}\,h^{(1)})\,A_{13}(\boldsymbol q)
\,A_{12}(\boldsymbol q-\eps_{R}\,h^{(3)})
\label{eq:dynYBE-a}\,,\\
D_{12}(\boldsymbol q+\eps_{L}\,h^{(3)})\,D_{13}(\boldsymbol q)
\,D_{23}(\boldsymbol q+\eps_{L}\,h^{(1)})
\, = \,D_{23}(\boldsymbol q)\,D_{13}(\boldsymbol q+\eps_{L}\,h^{(2)})\,
D_{12}(\boldsymbol q)
\label{eq:dynYBE-b}\,,\\
A_{12}(\boldsymbol q)\,C_{13}(\boldsymbol q-\eps_{R}\,h^{(2)})\,C_{23}(\boldsymbol q)
\, = \,C_{23}(\boldsymbol q-\eps_{R}\,h^{(1)})\,C_{13}(\boldsymbol q)
\,A_{12}(\boldsymbol q+\eps_{L}\,h^{(3)})
\label{eq:dynYBE-c}\,,\\
D_{12}(\boldsymbol q-\eps_{R}\,h^{(3)})\,B_{13}(\boldsymbol q)\,
B_{23}(\boldsymbol q+\eps_{L}\,
h^{(1)})
\, = \,B_{23}(\boldsymbol q)\,B_{13}(\boldsymbol q+\eps_{L}\,h^{(2)})\,
D_{12}(\boldsymbol q)
\label{eq:dynYBE-d}\,,
\eea
where the classical limit is defined by setting
\bea
R(\boldsymbol q) &=& \II +\hbar\,r(\boldsymbol q)+o(\hbar^2)\,,\quad 
R=A,B,C,D\mb{and}r=a,b,c,d
\\
h^{(i)} &=& \hbar\,h_{i}+o(\hbar^3)\,,
\eea
and keeping the order $\hbar^2$ in 
(\ref{eq:dynYBE-a})-(\ref{eq:dynYBE-d}), orders $1$ and $\hbar$ 
being trivial.

These 4 equations are in turn characterized as sufficient conditions 
for associativity of a quantum quadratic dynamical exchange algebra:
\beq
A_{12}(\boldsymbol q)\,K_{1}(\boldsymbol q-\eps_{R}\,h^{(2)})\,B_{12}(\boldsymbol q)\,
K_{2}(\boldsymbol q+\eps_{L}\,h^{(1)})
\, = \, K_{2}(\boldsymbol q-\eps_{R}\,h^{(1)})\,C_{12}(\boldsymbol q)\,
K_{1}(\boldsymbol q+\eps_{L}\,h^{(2)})\,D_{12}(\boldsymbol q)\,
\eeq
assuming a set of zero-weight conditions
\bea
&&\eps_{R}\,{[h^{(1)}+h^{(2)}\,,\,A_{12}]}\ =\  
\eps_{L}\,{[h^{(1)}+h^{(2)}\,,\,D_{12}]}\ =\ 0
\label{eq:zerowAD}\\
&&{[\eps_{R}\,h^{(1)}-\eps_{L}\,h^{(2)}\,,\,C_{12}]}\ =\  
{[\eps_{L}\,h^{(1)}-\eps_{R}\,h^{(2)}\,,\,B_{12}]}\ =\  0\,,
\label{eq:zerowBC}
\eea
and unitary hypothesis
\bea
&& A_{12}\,A_{21}\ =\ D_{12}\,D_{21}=\II\otimes\II\mb{;} C_{12}\ =\ B_{21}\,.
\eea
Altogether, these relations ensure associativity of the product in 
the dynamical algebra.\\

Note that it is always possible to redefine the overall sign of the 
dynamical variable $\boldsymbol q$, which in turn leads to a global sign 
change of $\eps_{R}$, $\eps_{L}$. It follows that only the relative 
sign between $\eps_{R}$ and  $\eps_{L}$ has relevance.

Note finally that the zero-weight conditions (\ref{eq:zerowBC}) put strong constraints on acceptable values
of the ratio of the c-numbers $\eps_{R}$, $\eps_{L}$ unless $B_{12}$ and $C_{12}$ belong to $\fh \otimes\fh$:
they must be ratios of weights of the corresponding Lie algebra.
We shall not discuss this issue any more at this time.

There exist in the literature two examples of the above algebra:
\begin{enumerate}
\item The dynamical boundary algebra \cite{BPOB} corresponds to $\eps_{L}=1$ 
and $\eps_{R}=-1$. We are not aware at this stage of 
explicit classical examples for the realization of 
(\ref{eq:rl-class}).
\item The semi-dynamical reflection algebra has 
$\eps_{R}=0$ and $\eps_{L}=1$. It is classically realized as 
(\ref{eq:dyn-class}) and (\ref{eq:rl-class}) by the Lax representation of the 
rational Ruijsenaars-Schneider model \cite{RS}.
\end{enumerate}
Our proposition corresponds to $\eps_{L}=\eps_{R}=1$. The dynamical 
Yang-Baxter equations now read:
\bea
A_{12}(\boldsymbol q)\,A_{13}(\boldsymbol q-h^{(2)})\,A_{23}(\boldsymbol q)
\, = \,A_{23}(\boldsymbol q-h^{(1)})\,A_{13}(\boldsymbol q)
\,A_{12}(\boldsymbol q-h^{(3)})
\label{eq:dynYBEnou-a}\,,\\
D_{12}(\boldsymbol q+h^{(3)})\,D_{13}(\boldsymbol q)
\,D_{23}(\boldsymbol q+h^{(1)})
\, = \,D_{23}(\boldsymbol q)\,D_{13}(\boldsymbol q+h^{(2)})\,D_{12}(\boldsymbol q)
\label{eq:dynYBEnou-b}\,,\\
A_{12}(\boldsymbol q)\,C_{13}(\boldsymbol q-h^{(2)})\,C_{23}(\boldsymbol q)
\, = \,C_{23}(\boldsymbol q-h^{(1)})\,C_{13}(\boldsymbol q)
\,A_{12}(\boldsymbol q+h^{(3)})
\label{eq:dynYBEnou-c}\,,\\
D_{12}(\boldsymbol q-h^{(3)})\,B_{13}(\boldsymbol q)\,B_{23}(\boldsymbol q+h^{(1)})
\, = \,B_{23}(\boldsymbol q)\,B_{13}(\boldsymbol q+h^{(2)})\,D_{12}(\boldsymbol q)
\label{eq:dynYBEnou-d}\,.
\eea
It is at least 
partially realized (at least for (\ref{eq:rl-class})) by the $2\times2$ Lax and 
$abcd$ matrices describing the second Poisson structure of the 
Calogero-Moser model \cite{AR2}. It is therefore justified to consider it for its 
own sake.

\section{Explicit realizations: the twisted dynamical Yangian\label{sec:expreal}}
We shall first of all establish the existence of this third dynamical 
reflection algebra and its associated dynamical Yang-Baxter equations 
by giving an explicit realization of these Yang-Baxter equations from 
any given dynamical $R$-matrix $A$ obeying (\ref{eq:dynYBE-a}). We 
recall that from now on we fix $\eps_{L}=\eps_{R}=1$.

Before proceeding, we must introduce notations and several technical 
lemmas which are essential to the manipulation of dynamical 
Yang-Baxter equations.
\begin{definition} Given a matrix 
\beq
M(\boldsymbol q)=\sum_{i,j=1}^n M_{ij}(\boldsymbol q)\, e_{ij}\in 
sl(n,\CC)\,,
\eeq
one denotes by respectively $M^{\texttt{\textit{sl}}}$ and $M^{\texttt{\textit{sc}}}$ the 
shifted matrices 
\bea
M^{\texttt{\textit{sl}}}(\boldsymbol q)=
\sum_{i,j=1}^n e^{\mu\,\prt_{q_i}}\,M_{ij}(\boldsymbol q)\,e^{-\mu\,\prt_{q_i}}\, e_{ij}
\in sl(n,\CC)\,,\\
M^{\texttt{\textit{sc}}}(\boldsymbol q)=
\sum_{i,j=1}^n e^{\mu\,\prt_{q_j}}\,M_{ij}(\boldsymbol q)\,e^{-\mu\,\prt_{q_j}}\, e_{ij}
\in sl(n,\CC)\,.
\eea
They are consistently defined as $\CC$-number matrices with shifts on 
$q_{k}$ coordinates in each matrix element $M_{ij}(\boldsymbol q)$, 
resp. $q_{k}\to q_{k}+\mu\,\delta_{ik}$ and $q_{k}\to q_{k}+\mu\,\delta_{jk}$.
In matrix form, the definition reads
\bea
M(\boldsymbol q)^{\texttt{\textit{sc}}}\ =\ 
\Big(e^{h\prt}\,\Big(M(\boldsymbol q)\,e^{-h\prt}\Big)^t\Big)^t
\mb{and}
M(\boldsymbol q)^{\texttt{\textit{sl}}}\ =\ 
\Big(\Big(e^{h\prt}\,M(\boldsymbol q)\Big)^t\,e^{-h\prt}\Big)^t
\,,
\eea
where $h\prt$ is defined in (\ref{eq:h-prt}).
\end{definition}
More generally, on a tensor product matrix one shall define  
shift $\texttt{\textit{sl}}_{a}$ and $\texttt{\textit{sc}}_{a}$ self-explanatorily.
\begin{lemma} If $M_{12}$ is zero-weight under adjoint action of 
$\fh\otimes\fh$ as
\beq
{[h^{(1)}+h^{(2)}\,,\,M_{12}]}=0
\eeq
then the coadjoint action of 
$\exp\sum_{j}\mu\,\big(h^{(1)}_{j}\prt_{q_j}+h^{(2)}_{j}\prt_{q_j}\big)$ on 
$M_{12}$ yields a $\CC$-number matrix
\beq
\wt M_{12} = M_{12}^{\texttt{\textit{sl}}_{1},\texttt{\textit{sc}}_{2}}
= M_{12}^{\texttt{\textit{sc}}_{1},\texttt{\textit{sl}}_{2}}\,.
\eeq
In other words, one has
\beq
e^{h^{(1)}\prt}\,M_{12}(\boldsymbol q)\,e^{-h^{(2)}\prt}=
e^{-h^{(2)}\prt}\,\wt M_{12}(\boldsymbol q)\, e^{h^{(1)}\prt}\,.
\eeq
\end{lemma}
The lemma is proven by direct calculation. We are now able to 
prove the following result
\begin{theorem}\label{theo:solBCD}
If $A_{12}(z_{1},z_{2};\boldsymbol q)$ is unitary and obeys the dynamical Yang-Baxter equation
 (\ref{eq:dynYBE-a}) with 
zero-weight condition $[h^{(1)}+h^{(2)}\,,\,A_{12}]=0$, we define the 
following matrices 
\bea
&&D_{12}(z_{1},z_{2};\boldsymbol q)=
\left(A_{21}^{t_{1}t_{2}}(f(z_{2}),f(z_{1});\boldsymbol q)\right)^{\texttt{\textit{sl}}_{1},\texttt{\textit{sl}}_{2}}
\,,\ \\ 
&&C_{12}(z_{1},z_{2};\boldsymbol q)=
\left(A_{21}^{t_{2}}(f(z_{2}),z_{1};\boldsymbol q)\right)^{\texttt{\textit{sc}}_{2}}
\\
&&B_{12}(z_{1},z_{2};\boldsymbol q)=
\left(A_{12}^{t_{1}}(f(z_{1}),z_{2};\boldsymbol q)\right)^{\texttt{\textit{sc}}_{1}}=C_{21}(z_{2},z_{1};\boldsymbol q)
\label{def:BCD}
\eea
where $f(z)$ is any $\CC$-valued function of the spectral parameter. 

These matrices
obey the Yang-Baxter equations 
(\ref{eq:dynYBE-b})-(\ref{eq:dynYBE-d}) and
$D_{12}(z_{1},z_{2};\boldsymbol q)$ is unitary.
\end{theorem}
The proof is again by direct calculation, using relations of the type
\beq
\big(M\,e^{h\prt}\,\wh P\big)^t = \left(\wh P^t\right)^{\texttt{\textit{sc}}}
\,e^{h\prt}\,\left( M^t\right)^{-\texttt{\textit{sl}}}\,,
\eeq
valid for any matrices $M$ and $\wh P$, and various partial 
transpositions of the Yang-Baxter equation for $A$.

\begin{rmk}
It is generally believed that dynamical $R$-matrices (obeying Gervais-Neveu-Felder equation) can be Drinfeld twisted to non dynamical $R$-matrices. The Drinfeld twist is constructed as solution of a master linear equation, now called the ABRR equation \cite{ABRR}.
In particular, Hecke-type dynamical $sl(n)$ $R$-matrices have been proved \cite{BRT2} to be (Drinfeld-) twisted to non-dynamical Cremmer-Gervais $R$-matrices\cite{CrGe}. The Drinfeld twist is in this case a dynamical coboundary \cite{BRT2}. 

Therefore, when the dynamical reflection algebras obey theorem \ref{theo:solBCD}, it is to be expected that they will be also twisted to non-dynamical reflection algebras. A similar result for the dynamical boundary algebra was presented in \cite{Hadj}.
\end{rmk}

\begin{theorem}\label{theo:tau}
We consider a representation of the dynamical Yang-Baxter equation for $A$ 
by a quantum Lax matrix $T$in $\mbox{End}(\cV)\otimes\mbox{End}(\cH)$ 
for $\cH$ a (quantum) Hilbert space, assuming in addition a 
zero-weight condition for $T$ under $\fh + \fh^{(q)}$. Here
$\cH$ too is a diagonalizable module of $\fh$ in order to be able to define $\fh^{(q)}$:
\bea
&&A_{12}(z_{1},z_{2};\boldsymbol q)\,T_{1}(z_{1};\boldsymbol q-h^{(2)})\,T_{2}(z_{2};\boldsymbol q) = 
T_{2}(z_{2};\boldsymbol q-h^{(1)})\,T_{1}(z_{1};\boldsymbol q)\,A_{12}(z_{1},z_{2};\boldsymbol q-h^{(q)})
\qquad\ \ \\
&&{[h^{(1)}+h^{(q)}\,,\, T_{1q}(z)]} =0\,,
\eea
where $h^{(q)}$ denotes the action of the Cartan algebra generators on the 
Hilbert space $\cH$, that is assumed to be a diagonalizable module of 
the Cartan algebra.

We also define the "transposed" Lax matrix $\cT$ as 
\beq
\cT(z;\boldsymbol q)= \left(T^{t}(f(z);\boldsymbol q)\right)^{\texttt{\textit{sc}}}\,,
\eeq
where $f(z)$ is the same function as in Theorem \ref{theo:solBCD}. Transposition and shifts here act on the
auxiliary space indices.
It obeys a transposed exchange relation and a crossed exchange 
relation:
\bea
&&D_{12}(z_{1},z_{2};\boldsymbol q-h^{(q)})\, \cT_{1}(z_{1};\boldsymbol q)\,\cT_{2}(z_{2};\boldsymbol q+h^{(1)})
=
\cT_{2}(z_{2};\boldsymbol q)\, \cT_{1}(z_{1};\boldsymbol q+h^{(2)})\,D_{12}(z_{1},z_{2};\boldsymbol q)
\ \qquad\ \\
&&\cT_{1}(z_{1};\boldsymbol q-h^{(2)})\,B_{12}(z_{1},z_{2};\boldsymbol q)\,T_{2}(z_{2};\boldsymbol q+h^{(1)})
=
T_{2}(z_{2};\boldsymbol q)\,C_{21}(z_{1},z_{2};\boldsymbol q-h^{(q)})\,\cT_{1}(z_{1};\boldsymbol q)\ 
\qquad\ 
\eea
where the $B,C,D$ matrices are given as in (\ref{def:BCD}). If a 
$\CC$-number matrix $\gamma$ exists such that
\beq
A_{12}(z_{1},z_{2};\boldsymbol q)\,\gamma_{1}(z_{1};\boldsymbol q)\,B_{12}(z_{1},z_{2};\boldsymbol q)\,
\gamma_{2}(z_{2};\boldsymbol q)= 
\gamma_{2}(z_{2};\boldsymbol q)\,C_{12}(z_{1},z_{2};\boldsymbol q)\,\gamma_{1}(z_{1};\boldsymbol q)
\,D_{12}(z_{1},z_{2};\boldsymbol q)
\eeq
then the operator valued matrix 
\beq
K(z;\boldsymbol q)=T(z;\boldsymbol q)\,\gamma(z;\boldsymbol q+h^{(q)})\,\cT(z;\boldsymbol q)
\eeq
realizes the 
dynamical quadratic algebra (\ref{eq:ABCDdyn3}).
\end{theorem}
Again, the proof is done by direct calculation.

Examples of $\gamma$ matrices for the dynamical twisted reflection algebra 
will be given in section \ref{sec:ex}, thereby establishing the existence
of explicit realizations of the proposed third dynamical reflection algebra.
\begin{rmk}
This construction can be characterized as a dynamical extension of the 
twisted (quantum) Yangian construction \cite{olsh,MNO,qYangtw}, in the specific 
case when the anti-automorphism $\sigma$ is chosen to be the 
transposition. 
\end{rmk}
\begin{rmk}
A similar construction occurs for the boundary dynamical algebra 
($\eps_{L}=1=-\eps_{R}$) except that $\sigma$ is there chosen to be the 
inverse. Modifications on the definition of $B,C,D$ from $A$ imply 
that one has $A=C=B^{-1}=D^{-1}$. Hence, in the boundary dynamical 
algebra case, the existence of at least one matrix $\gamma=\II$ is trivially
guaranteed.

\end{rmk}

In the generic $ABCD$ case, one proves the existence of two coactions:
\begin{theorem}\label{theo:dressK}
Let us assume that $K$ obeys the dynamical exchange algebra
\beq
A_{12}(\boldsymbol q)\,K_{1}(\boldsymbol q-h^{(2)})\,B_{12}(\boldsymbol q)\,K_{2}(\boldsymbol q+h^{(1)})
=
K_{2}(\boldsymbol q-h^{(1)})\,C_{12}(\boldsymbol q)\,K_{1}(\boldsymbol q+h^{(2)})\,D_{12}(\boldsymbol q)
\label{eq:dynK}
\eeq
where $A,B,C,D$ obey (\ref{eq:dynYBEnou-a})-(\ref{eq:dynYBEnou-d}).
Introduce $L$ and $J$ obeying respectively the following exchange relations
\bea
&& A_{12}(\boldsymbol q)\,L_{1}(\boldsymbol q-h^{(2)})\,L_{2}(\boldsymbol q)
=L_{2}(\boldsymbol q-h^{(1)})\,L_{1}(\boldsymbol q)\,A_{12}(\boldsymbol q+\alpha\,h^{(q)})
\label{eq:th3-a}\\
&& D_{12}(\boldsymbol q+\alpha\,h^{(q)})\,J_{1}(\boldsymbol q)\,J_{2}(\boldsymbol q+h^{(1)})
=J_{2}(\boldsymbol q)\,J_{1}(\boldsymbol q+h^{(2)})\,D_{12}(\boldsymbol q)
\label{eq:th3-b}\\
&& J_{1}(\boldsymbol q-h^{(2)})\,B_{12}(\boldsymbol q)\,L_{2}(\boldsymbol q+h^{(1)})
=L_{2}(\boldsymbol q)\,B_{12}(\boldsymbol q+\alpha\,h^{(q)})\,J_{1}(\boldsymbol q)
\label{eq:th3-c}\\
&& J_{2}(\boldsymbol q-h^{(1)})\,C_{12}(\boldsymbol q)\,L_{1}(\boldsymbol q+h^{(2)})
=L_{1}(\boldsymbol q)\,C_{12}(\boldsymbol q+\alpha\,h^{(q)})\,J_{2}(\boldsymbol q)
\label{eq:th3-d}\\
&&{[\alpha\,h^{(q)}-h^{(1)}\,,\, L_{1}]}=0\mb{and}
{[\alpha\,h^{(q)}+h^{(1)}\,,\, J_{1}]}=0
\label{eq:th3-e}
\eea
where $q$ labels a Hilbert space $\cH$ carried by $T$ and $L$ and 
$\alpha$ is any $\CC$-number. Then,
\beq
\wt K(\boldsymbol q)=L(\boldsymbol q)\,K(\boldsymbol q+\alpha\,h^{(q)})\,T(\boldsymbol q)
\label{eq:dressK}
\eeq
realizes also (\ref{eq:dynK}).
\end{theorem}
The proof follows again from a direct computation.
In the reduced situation described by Theorem \ref{theo:solBCD}, an example of the 
construction described in Theorem \ref{theo:dressK} is precisely 
given by Theorem \ref{theo:tau}.
\paragraph{Example 1: $\boldsymbol{\alpha=1}$}
Taking $J_{1}(z_{1})=D_{1a}(z_{1},0)$ and $L_{1}(z_{1})=C_{1a}(z_{1},0)$ realize (\ref{eq:th3-a})-(\ref{eq:th3-e}). The 
quantum space $\cH$ is then identified with the auxiliary space 
$\cV$, as it is standard in the construction of spin chain monodromy 
matrices.
\paragraph{Example 2: $\boldsymbol{\alpha=-1}$}
Taking $J_{1}(z_{1})=B_{1a}(z_{1},0)$ and $L_{1}(z_{1})=A_{1a}(z_{1},0)$ realize (\ref{eq:th3-a})-(\ref{eq:th3-e}).
Again, in the reduced situation described by Theorem \ref{theo:solBCD},
successive implementation of this coaction realizes precisely 
(\ref{eq:dressK}), with
\beq
T_{1}(z_{1};\boldsymbol q)=\prod_{1\leq i\leq n}^{\longleftarrow} 
A_{1,a_{i}}\Big(z_1,0;\boldsymbol q-\sum_{j=i+1}^nh^{(a_j)}\Big)\,,
\eeq
a well-known formula for the homogeneous dynamical bulk monodromy, 
see \cite{Fe,ABB}.

These two examples can be combined to build realizations of the 
dynamical exchange algebra as dressing of an initial scalar 
solution $\gamma$ (to be computed) by successive pairs $(D,C)$ and 
$(B,A)$ resp. on the right and left side of the $(n-1)$ sites 
monodromy matrix, together with consistent shifts on $\boldsymbol q$ of the ``internal'' 
generators of this  monodromy matrix according to (\ref{eq:dressK}).

\section{Commuting traces\label{sec:commuting}}
The general procedure to build a generating functional for commuting 
operators associated with the algebraic structure (\ref{eq:dynK}) follows 
from arguments formally similar to the cases of the other two dynamical reflection algebras. It is 
summarized in the following theorem
\begin{theorem}\label{theo:dual}
The following operators
\beq
H_{j}=\mbox{Tr}_{j}\Big(\,e^{\prt_{j}}\,K_{j}(z;\boldsymbol q)\,e^{\prt_{j}}
\,K^+_{j}(z;\boldsymbol q)^t\Big)
\label{eq:Ham}
\eeq
commute with one another for every choice of pairs of distinct 
auxiliary spaces $\cV_{j}, \cV_{k}$. The trace Tr$_{j}$ is \underline{only} 
taken over the vector space indices whenever $\cV_{j}$ is a loop 
space $\cV_{j}=V_{j}\otimes\CC[[z_{j}]]$. The notation $\prt_{j}$ is 
shorthand for
\beq
\prt_{j}=\sum_{k=1}^n \mu\,h^{(j)}_{kk}\,\frac{\prt}{\prt q_{k}}\,.
\eeq
$K_{(j,k)}(z;\boldsymbol q)$ obey the general exchange relation (\ref{eq:dynK}) for distinct auxiliary
spaces $\cV_{j}, \cV_{k}$, and 
 $K_{(j,k)}^+$ obey the dual 
dynamical reflection equation
\beq\label{eq:dynKdual}
\wt A_{jk}(\boldsymbol q)\,
K^+_{j}(\boldsymbol q-h^{(k)})^{\texttt{\textit{sc}}_{j}}\,
\wt B_{jk}(\boldsymbol q)\,
K^+_{k}(\boldsymbol q+h^{(j)})^{\texttt{\textit{sc}}_{k}}
\ =\ 
K^+_{k}(\boldsymbol q-h^{(j)})^{\texttt{\textit{sc}}_{k}}\,
\wt C_{jk}(\boldsymbol q)\,
K^+_{j}(\boldsymbol q+h^{(k)})^{\texttt{\textit{sc}}_{j}}\,
\wt D_{jk}(\boldsymbol q)
\qquad
\eeq
where 
\bea
\wt A_{jk}(\boldsymbol q)=\Big(A_{jk}^{-1}(\boldsymbol q)\Big)^{t_{j}t_{k}}
\mb{;}
\wt B_{jk}(\boldsymbol q)=\Big(\big(B_{jk}^{t_{k}}(\boldsymbol q)\big)^{-1}\Big)^{t_{j}}\,
\\
\wt C_{jk}(\boldsymbol q)=
\Big(\big(C_{jk}^{t_{j}}(\boldsymbol q)\big)^{-1}\Big)^{t_{k}}
=\wt B_{kj}(z_{k},z_{j};\boldsymbol q)
\mb{;}
\wt D_{jk}(\boldsymbol q)=\Big(D_{jk}^{t_{j}t_{k}}(\boldsymbol q)\Big)^{-1}
\eea
\end{theorem}
The proof of this theorem is a long and technical calculation that 
has been detailed in \cite{thZ} for another type of dynamical reflection algebras. 
In our case, the proof follows the same lines, with appropriate change 
of signs.

Such a choice of distinct auxiliary spaces leading to non-trivial sets of commuting
quantum Hamiltonians is available in at least two well-known situations:

First of all when the auxiliary spaces $\cV_{j}, \cV_{k}$ are isomorphic (but not identical) loop spaces, the restricted 
trace over vector indices
yields a generating function $t(z)$ where $z$ is the spectral parameter. The Theorem then establishes that
$[t(z_1), t(z_2) = 0$. Hence the operatorial coefficients of the formal series expansion of $t(z)$ in powers
of $z$ provide a set of mutually commuting Hamitonians. This is the standard procedure in e.g. the case of quantum
integrable spin chains.

The second case corrsponds to the so-called quantum power traces such as originally described in \cite{Maillet}. It stems 
from the existence of a systematic procedure to construct successive tensorial powers of an initial finite
dimensional auxiliary space together with the corresponding coefficient matrices $ABCD$. This procedure will be described
presently in Sections 5 and 6 as ``fusion'' and ``dressing''.

Let us now discuss more precisely the dual reflection equation.
One can define a reduced representation of the coefficient matrices �
of the dual equation following Theorem \ref{theo:solBCD}:
\begin{corollary}
If $A,B,C,D$ obey the relations given in theorem \ref{theo:solBCD}
\bea
&&D_{12}(z_{1},z_{2};\boldsymbol q)=
\left(A_{21}^{t_{1}t_{2}}(f(z_{2}),f(z_{1});\boldsymbol q)\right)^{\texttt{\textit{sl}}_{1},\texttt{\textit{sl}}_{2}}
\,,\ \\ 
&&C_{12}(z_{1},z_{2};\boldsymbol q)=
\left(A_{21}^{t_{2}}(f(z_{2}),z_{1};\boldsymbol q)\right)^{\texttt{\textit{sc}}_{2}}
\\
&&B_{12}(z_{1},z_{2};\boldsymbol q)=
\left(A_{12}^{t_{1}}(f(z_{1}),z_{2};\boldsymbol q)\right)^{\texttt{\textit{sc}}_{1}}=C_{21}(z_{2},z_{1};\boldsymbol q)
\eea
then the same relations are valid for $\wt A,\wt B,\wt C,\wt D$ given 
in theorem \ref{theo:dual}:
\bea
&&\wt D_{12}(z_{1},z_{2};\boldsymbol q)=
\left(\wt A_{21}^{t_{1}t_{2}}(f(z_{2}),f(z_{1});\boldsymbol q)\right)^{\texttt{\textit{sl}}_{1},\texttt{\textit{sl}}_{2}}
\,,\ \\ 
&&\wt C_{12}(z_{1},z_{2};\boldsymbol q)=
\left(\wt A_{21}^{t_{2}}(f(z_{2}),z_{1};\boldsymbol q)\right)^{\texttt{\textit{sc}}_{2}}
\\
&&\wt B_{12}(z_{1},z_{2};\boldsymbol q)=
\left(\wt A_{12}^{t_{1}}(f(z_{1}),z_{2};\boldsymbol q)\right)^{\texttt{\textit{sc}}_{1}}
\eea
\end{corollary}

In addition, this reduced representation induces a relation between direct
and dual scalar solutions as follows
\begin{corollary}
If $A,B,C,D$ obey the relations given in theorem \ref{theo:solBCD} 
with $f(z)=-z$, and if moreover $A$ obeys the crossing relation
\beq
\left(A_{12}^{t_{1}}(z_{1},z_{2};\boldsymbol q)\right)^{-1}= 
\left(A_{12}^{-1}(z_{1}+\frac\eta2,z_{2}-\frac\eta2;\boldsymbol q)\right)^{t_{2}}
\eeq
then, from any solution $\KK(z;\boldsymbol q)$ to the dynamical 
reflection equation (\ref{eq:dynK}), one can construct a solution 
$K^{+}(z;\boldsymbol q)$ to the dual dynamical 
reflection equation (\ref{eq:dynKdual}) as
\beq
K^{+}(z;\boldsymbol q)=\left(\left(\KK^t(z+\frac\eta2;\boldsymbol 
q)\right)^{-1}\right)^{-\texttt{\textit{sc}}}
\eeq
In that case, the expression of Hamiltonian (\ref{eq:Ham}) simplifies 
to
\beq
H_{j}=\mbox{Tr}_{j}\Big(\,e^{\prt_{j}}\,K_{j}(z;\boldsymbol q)
\,\KK_{j}(z;\boldsymbol q)^{-1}\,e^{\prt_{j}}\Big)
\label{eq:Ham2}
\eeq
with now $K_{j}$ and $\KK_{j}$ solutions to the same dynamical 
reflection equation (\ref{eq:dynK}).
\end{corollary}

\section{Fusion procedures\label{sec:fusion}}
The coaction procedure in Section 3 described tensoring 
of "quantum" spaces. We now turn to the fusion procedure, that allows 
to construct higher spin representations of the quantum algebra by a 
consistent tensoring of auxiliary spaces. This "auxiliary" tensoring 
also play a key role (as commented before) in defining higher powers in quantum traces of 
monodromy matrices, realizing through application
of the quantum trace formula the quantum analogue of the classical 
$Tr(L^n)$ for a Lax matrix (see \cite{Maillet,ABB,NADR}).

We shall restrict ourselves to the first step, i.e. definition of a 
consistent tensor product on auxiliary spaces. The next step, i.e. 
projection on irreducible representations, is a complex and delicate 
issue which should be examined separately.

The tensoring procedure actually follows from two fundamental lemmas, 
proved by direct computation using suitable Yang-Baxter equations:
\begin{lemma}
\label{lem:fus}
Given two representations of the dynamical exchange algebra, 
respectively:
\bea
&&A_{12}(\boldsymbol q)\,K_{1}(\boldsymbol q-h^{(2)})\,B_{12}(\boldsymbol q)\,K_{2}(\boldsymbol q+h^{(1)})
=
K_{2}(\boldsymbol q-h^{(1)})\,C_{12}(\boldsymbol q)\,K_{1}(\boldsymbol q+h^{(2)})\,D_{12}(\boldsymbol q)
\qquad\quad\\
&&A_{12}(\boldsymbol q)\,K'_{1}(\boldsymbol q-h^{(2)})\,B_{12}(\boldsymbol q)\,K'_{2}(\boldsymbol q+h^{(1)})
=
K'_{2}(\boldsymbol q-h^{(1)})\,C_{12}(\boldsymbol q)\,K'_{1}(\boldsymbol q+h^{(2)})\,D_{12}(\boldsymbol q)
\qquad\quad
\eea
with the same quantum space $\cH$, the following $K$-matrices and 
structural matrices:
\bea
&&A_{<11'>2}(\boldsymbol q)=A_{1'2}(\boldsymbol q-h^{(1)})\,A_{12}(\boldsymbol q)\mb{;}
D_{<11'>2}(\boldsymbol q)=D_{1'2}(\boldsymbol q)\,D_{12}(\boldsymbol q+h^{(1')})\mb{;}
\\
&&B_{<11'>2}(\boldsymbol q)=B_{1'2}(\boldsymbol q)\,B_{12}(\boldsymbol q+h^{(1')})\mb{;}
C_{<11'>2}(\boldsymbol q)=C_{1'2}(\boldsymbol q-h^{(1)})\,C_{12}(\boldsymbol q)\mb{;}
\\
&&h^{(<11'>)}=h^{(1)}+h^{(1')}\mb{and}
K_{<11'>}(\boldsymbol q)=K'_{1'}(\boldsymbol q-h^{(1)})\,B_{11'}(\boldsymbol q)\,,K_{1}(\boldsymbol q+h^{(1')})
\qquad
\eea
obey the dynamical exchange algebra:
\bea
&&A_{<11'>2}(\boldsymbol q)\,K_{<11'>}(\boldsymbol q-h^{(2)})\,B_{<11'>2}(\boldsymbol q)
\,K_{2}(\boldsymbol q+h^{(<11'>)})=
\nonu
&&\qquad=K_{2}(\boldsymbol q-h^{(<11'>)})\,C_{<11'>2}(\boldsymbol q)\,K_{<11'>}(\boldsymbol q+h^{(2)})\,D_{12}(\boldsymbol q)
\eea
\end{lemma}
\begin{lemma}
\label{lem:fus2}
Given two representations of the dynamical exchange algebra, 
respectively:
\bea
&&A_{12}(\boldsymbol q)\,K_{1}(\boldsymbol q-h^{(2)})\,B_{12}(\boldsymbol q)\,K_{2}(\boldsymbol q+h^{(1)})
=
K_{2}(\boldsymbol q-h^{(1)})\,C_{12}(\boldsymbol q)\,K_{1}(\boldsymbol q+h^{(2)})\,D_{12}(\boldsymbol q)
\qquad\\
&&A_{12}(\boldsymbol q)\,K'_{1}(\boldsymbol q-h^{(2)})\,B_{12}(\boldsymbol q)\,K'_{2}(\boldsymbol q+h^{(1)})
=
K'_{2}(\boldsymbol q-h^{(1)})\,C_{12}(\boldsymbol q)\,K'_{1}(\boldsymbol q+h^{(2)})\,D_{12}(\boldsymbol q)
\qquad
\eea
with the same quantum space $\cH$, the following $K$-matrices and 
structural matrices:
\bea
&&A_{1<22'>}(\boldsymbol q)=A_{12}(\boldsymbol q)\,A_{12'}(\boldsymbol q-h^{(2)})\mb{;}
D_{1<22'>}(\boldsymbol q)=D_{12}(\boldsymbol q+h^{(2')})\,D_{12'}(\boldsymbol q)\mb{;}
\\
&&B_{1<22'>}(\boldsymbol q)=B_{12'}(\boldsymbol q-h^{(2')})\,B_{12}(\boldsymbol q)\mb{;}
C_{1<22'>}(\boldsymbol q)=C_{12'}(\boldsymbol q)\,C_{12}(\boldsymbol q-h^{(2')})\mb{;}
\\
&&h^{(<22'>)}=h^{(2)}+h^{(2')}\mb{and}
K_{<22'>}(\boldsymbol q)=K'_{2'}(\boldsymbol q-h^{(2)})\,B_{2'2}(\boldsymbol q)\,,K_{2}(\boldsymbol q+h^{(2')})
\qquad
\eea
obey the dynamical exchange algebra:
\bea
&&A_{1<22'>}(\boldsymbol q)\,K_{1}(\boldsymbol q-h^{(<22'>)})\,B_{1<22'>}(\boldsymbol q)
\,K_{<22'>}(\boldsymbol q+h^{(1)})=
\nonu
&&\qquad=
K_{<22'>}(\boldsymbol q-h^{(1)})\,C_{1<22'>}(\boldsymbol q)\,K_{1}(\boldsymbol q+h^{(<22'>)})\,
D_{1<22'>}(\boldsymbol q)
\eea
\end{lemma}
Note that if $\cV_{1}=\cV_{2}$ and $\cV_{1'}=\cV_{2'}$, one can show 
that if one assumes the unitary relations
$C_{12}=B_{21}$, $A_{12}\,A_{21}=\II=D_{12}\,D_{21}$ and zero-weight 
condition for $A$ and $D$, one has
\bea
&&A_{<11'><22'>}\,A_{<22'><11'>}=\II=D_{<11'><22'>}\,D_{<22'><11'>}\,,
\\
&&C_{<11'><22'>}=B_{<22'><11'>}\,.
\eea
Hence, the unitary condition is indeed preserved whenever relevant.

Successive iterations of both Lemmae allow to define multiply fused
coefficient matrices $A_{\underline{M}\,\underline{N}}, B_{\underline{M}\,\underline{N}},
C_{\underline{M}\,\underline{N}}, D_{\underline{M}\,\underline{N}}$ and operator �
matrices $K_{\underline{M}}$ labeled by ordered sets of auxiliary space indices
$\underline{M}=<12\ldots m>$, $\underline{N}=<1'2'\ldots n'>$. �
Precisely, the fusion formulae are to be understood as describing the �
procedure of addition
of a single space index, resp. $1'$ to an already fused multiple space index
$1$ (left-hand fusion) and $2'$ to an already fused multiple space index
$2$ (right-hand fusion). The final formulae are quite cumbersome but closely
match the formulae given in Section IV-A of \cite{NADR} with suitable �
changes in the sign shifts.

It is important to remark here that the Yang-Baxter equations
guarantee that the fused coefficient matrices are univocally defined �
whichever order of implementation is defined to add left and right �
indices to resp. $\underline{M}$
and $\underline{N}$. This is not surprising since Yang-Baxter equations are originally
a guarantee of invariance under permutation of space indices in �
exchange processes.

\section{Dressing procedures\label{sec:dressing}}
As in the previous cases of (dynamical) boundary algebras, there 
exists a supplementary possibility to implement the dynamical 
quadratic algebra on a tensor product of auxiliary spaces by the 
so-called "dressing" procedure. Precisely, 
one has
\begin{lemma}
If $K$  obeys the general exchange relation (\ref{eq:dynK}), then the 
operator $Q\,K\,S$ also realizes this algebra, provided the 
operators $Q$ and $S$ verify the following relations
\bea
&&A_{\underline{N}\,\underline{M}}(\boldsymbol q)\,Q_{\underline{N}}(\boldsymbol q-h^{(\underline{M})}) 
= Q_{\underline{N}}(\boldsymbol q)\,A_{\underline{N}\,\underline{M}}(\boldsymbol q) 
\mb{with} h^{(\underline{M})}=h^{(1)}+h^{(2)}+\ldots+h^{(m)}
\qquad\label{eq:AQcom}\\
&&C_{\underline{N}\,\underline{M}}(\boldsymbol q)\,Q_{\underline{N}}(\boldsymbol q+h^{(\underline{M})}) 
= Q_{\underline{N}}(\boldsymbol q)\,C_{\underline{N}\,\underline{M}}(\boldsymbol q) 
\qquad\label{eq:CQcom}\\
&&D_{\underline{N}\,\underline{M}}(\boldsymbol q)\,S_{\underline{N}}(\boldsymbol q)
 = S_{\underline{N}}(\boldsymbol q+h^{(\underline{M})})\,D_{\underline{N}\,\underline{M}}(\boldsymbol q)
 \label{eq:DScom}\\
&&B_{\underline{N}\,\underline{M}}(\boldsymbol q)\,S_{\underline{N}}(\boldsymbol q)
 = S_{\underline{N}}(\boldsymbol q-h^{(\underline{M})})\,B_{\underline{N}\,\underline{M}}(\boldsymbol q)
 \label{eq:BScom}\\
&&{[h_{\underline{N}}\,,\,Q_{\underline{N}}]}=0={[h_{\underline{N}}\,,\,S_{\underline{N}}]} 
\label{eq:QSh}
\eea
\end{lemma}
It is to be understood in this Lemma that the auxiliary space indices �
$\underline{M},\underline{N},\ldots$ of the coefficient matrices are generically sets
of fused indices obtained by implementation of the 
fusion Lemmae \ref{lem:fus} and \ref{lem:fus2}.

An explicit realization of $Q$ and $S$ is obtained as follows:
\begin{lemma}
The following objects
\bea
&&
Q_{\underline{N}}\equiv Q_{1\ldots n}(\boldsymbol q)\ =\ 
\wchk{A}_{21}(\boldsymbol q)\,
\wchk A_{32}\big(\boldsymbol q+h^{(1)}\big)\,\cdots 
\wchk A_{n,n-1}\big(\boldsymbol q+h^{(1)}+\ldots+h^{(n-2)}\big)
\qquad 
\\
&&
S_{\underline{N}}\equiv S_{1\ldots n}(\boldsymbol q)\ =\ 
\wchk D_{21}\big(\boldsymbol q+h^{(3)}+\ldots+h^{(n)}\big)
\wchk D_{32}\big(\boldsymbol q+h^{(4)}+\ldots+h^{(n)}\big)\,\cdots 
\wchk D_{n,n-1}(\boldsymbol q)\qquad\ 
\eea
realize relations (\ref{eq:AQcom})-(\ref{eq:QSh}). We denote here 
$\wchk R_{12}=\cP_{12}\,R_{12}$, $\forall\,R$,
where $\cP_{ij}$ is the permutation operator acting on the auxiliary 
spaces $\cV_{i}\otimes \cV_{j}$.
\end{lemma}

\begin{rmk}
The dressing procedure yields commuting traces that are completely 
different from the ones obtained from the fusion procedure, as 
described in section \ref{sec:fusion}. In fact, one can show (at least 
in some particular cases) that the latter leads to completely 
factorized traces 
$\Big(\mbox{Tr}\,(K\,e^{h\prt}\wt K\,e^{h\prt})\Big)^n$, 
while the former is closer to a form
 $\mbox{Tr}\,\Big((K\,e^{h\prt}\wt K\,e^{h\prt})^n\Big)$. 
The dressing is useful for constructing an 
adequate set of independent quantum quantities, while fusion is more 
appropriate to built a local Hamiltonian.
This key point of the dressing procedure was already pointed out in 
\cite{ABB,Maillet,NADR}.
\end{rmk}

\begin{rmk}
The permutation operator $\cP_{12}$ acts of course as $\cP_{12}R_{12}\cP_{12} =
R_{21}$. If $R_{12}$ does not depend on spectral parameters (i.e. the
auxiliary spaces $V_1=V_2$ are isomorphic to finite dimensional diagonalizable
modules of the Cartan algebra $h$), the operator $P_{12}$
is easily constructed as $P = \sum_{i,j=1}^{n} E_{ij} \otimes E_{ji}$
where $E_{ij}$ are the elementary matrices of $End(V_{1})$.

If however $R_{12}$ depends on spectral parameters $z_1, z_2$ (i.e. the
auxiliary spaces $\cV_1$, $\cV_2$ are loop spaces $V\otimes\CC[[z_1]]$,
$V\otimes\CC[[z_2]]$, explicit implementation of the spectral parameter
exchange as $\cP_{12} f(z_1, z_2) \cP_{12} = f(z_2, z_1)$ is not so easily
available. It may then be difficult to explicitly define the quantum
commuting operators analogous to $\mbox{Tr}\,\Big((K\,e^{h\prt}\wt
K\,e^{h\prt})^n\Big)$. Fortunately, expansion in formal powers of the
spectral parameter $z$ of the single, well-defined
operator $\mbox{Tr}\,\Big(K(z)\,e^{h\prt}\wt K(z)\,e^{h\prt}\Big)$ is 
in this case available. It
provides an alternative viable procedure to obtain a family of
algebraically independent quantum commuting operators.
\end{rmk}

\section{Examples \label{sec:ex}}
We consider the following $A,B,C,D$ matrices acting in $\CC^n\otimes 
\CC^n$:
\bea
A_{12}&=& \II_{n}\otimes\II_{n} + \sum_{1\leq i\neq j\leq n} \
\frac{\mu}{q_{i}-q_{j}}\,\Big(E_{ij}\otimes E_{ji}-
E_{ii}\otimes E_{jj}\Big)
\\
B_{12}&=& \II_{n}\otimes\II_{n} + \sum_{1\leq i\neq j\leq n} \
\frac{\mu}{q_{i}-q_{j}+\mu}\,\Big(E_{ji}\otimes E_{ji}-
E_{ii}\otimes E_{jj}\Big)
\\
C_{12}&=& \II_{n}\otimes\II_{n} + \sum_{1\leq i\neq j\leq n} \
\frac{\mu}{q_{i}-q_{j}+\mu}\,\Big(E_{ji}\otimes E_{ji}-
E_{jj}\otimes E_{ii}\Big)
\\
D_{12}&=& \II_{n}\otimes\II_{n} + \sum_{1\leq i\neq j\leq n} \
\frac{\mu}{q_{i}-q_{j}}\,\Big(E_{ij}\otimes E_{ji}-
E_{ii}\otimes E_{jj}\Big)
\eea
where $E_{ij}$ is the $n\times n$ elementary matrix with 1 at 
position $(i,j)$ and 0 elsewhere. Those matrices are of the type 
given in Theorem \ref{theo:solBCD}.

Direct computations show that the following matrices 
\bea
\gamma &=& \sum_{1\leq i, j\leq n} 
\,m_{i}\,m_{j}\,E_{ij}\,,\qquad m_{i}\in\CC
\label{eq:sol1}\\
\gamma &=& \sum_{1\leq i, j\leq n} 
(q_{i}-q_{j})\,m_{i}\,m_{j}\,E_{ij}\,,\qquad m_{i}\in\CC
\label{eq:sol2}\\
\gamma &=& \sum_{i=1}^n 
\frac{f(\boldsymbol q+\mu\,e_{ii})}{f(\boldsymbol q-\mu\,e_{ii})}\ 
\prod_{k\neq i}(q_{i}-q_{k})\ E_{ii}\,,
\label{eq:sol3}
\eea
where $f$ is an arbitrary $\CC$-valued function and $e_{ii}\in\fh$,
obey the dynamical reflection equation. Let us note that the two first 
solutions are not invertible.

Then, expression (\ref{eq:Ham2}) with $K_{j}$ of the form 
(\ref{eq:sol1}) and 
$\KK_{j}$ of the form (\ref{eq:sol3}) leads to the Hamiltonian
\beq
H=\sum_{\ell=1}^n \frac{\mu\,m_{\ell}^2}{\prod_{k=1}^n
(q_{\ell}-q_{k}+\mu)}\,e^{2\mu\,\prt_{q_\ell}}\,.
\label{eq:Ham-ex}
\eeq
The $f$-dependent factor in (\ref{eq:sol3}) is easily seen to be 
reabsorbed by a conjugation of $H$ by $f(\boldsymbol q)$. It has been 
therefore set to 1 in (\ref{eq:Ham-ex}).
From our construction of sections \ref{sec:commuting}, \ref{sec:fusion} 
and \ref{sec:dressing}, these Hamiltonians are in principle quantum 
integrable. We remark that they take a form close to the 
Ruijsenaars-Schneider Hamiltonians, although the coinciding points 
singularities have been replaced by finite distance singularities.

For the particular case $n=2$, one gets
\beq 
H=\left(\frac{m_{1}^2}{q+\mu}\,e^{2\mu\,\prt_{q}}-\frac{m_{2}^2}{q-\mu}\,
e^{-2\mu\,\prt_{q}}\right)
e^{2\mu\,\prt_{Q}}
\eeq
where we have introduced the relative and center-of-mass coordinates
$q=q_{1}-q_{2}$ and $Q=q_{1}+q_{2}$.
The relative Hamiltonian has in particular eigenfunctions with 
zero eigenvalue taking the form
\beq
\begin{array}{l}
\displaystyle
\psi_{k}(q) = \frac{\Gamma(\frac{q+\mu}{4\mu}+\half)}{\Gamma(\frac{q+\mu}{4\mu})}
\,e^{-\frac q{4\mu}\, \ln(\frac{m_{1}}{m_{2}})}\,
\sin(k\,\frac{\pi q}{\mu})
\\
\displaystyle
\vph_{k}(q) = \frac{\Gamma(\frac{q+\mu}{4\mu}+\half)}{\Gamma(\frac{q+\mu}{4\mu})}
\,e^{-\frac q{4\mu}\, \ln(\frac{m_{1}}{m_{2}})}\,
\cos(k\,\frac{\pi q}{\mu})\end{array}
\qquad k\in\ZZ_{\geq0}\,.
\eeq

\subsection*{Acknowledgements}
This work was sponsored by CNRS, Universit\'e de Cergy-Pontoise, 
Universit\'e de Savoie and ANR Project 
DIADEMS (Programme Blanc ANR SIMI1 2010-BLAN-0120-02). J.A. wishes to 
thank LAPTh for their kind hospitality.


\end{document}